\documentclass{article}
\usepackage{ijcai22}

\usepackage{times}
\usepackage{soul}
\usepackage{url}
\usepackage[hidelinks,colorlinks,citecolor=blue]{hyperref}
\usepackage[utf8]{inputenc}
\usepackage[small]{caption}
\usepackage{graphicx}
\usepackage{amsmath}
\usepackage{amsthm}
\usepackage{booktabs}
\usepackage{algorithm}
\usepackage{algorithmic}
\usepackage{amsmath}
\usepackage{bm}
\usepackage{multirow}
\usepackage[table,xcdraw]{xcolor}
\usepackage{tabu}
\usepackage{graphicx}
\usepackage{amsmath}
\usepackage{amssymb}
\usepackage{booktabs}
\usepackage{makecell}
\usepackage{mathrsfs}
\usepackage{siunitx}
\usepackage{setspace}

\title{Robust Audio-Visual Instance Discrimination via Active Contrastive Set Mining}

\author{
Hanyu Xuan$^1$\and
Yihong Xu$^2$\and
Shuo Chen$^3$\and 
Zhiliang Wu$^1$\and \\
Jian Yang$^1$\and 
Yan Yan$^4$\thanks{Corresponding author}\and
Xavier Alameda-Pineda$^2$\\
\affiliations
$^1$School of Computer Science and Engineering, Nanjing University of Science and Technology, China\\
$^2$Inria, University Grenoble Alpes, CNRS, Grenoble INP, LJK, Grenoble, France\\
$^3$RIKEN Center for Advanced Intelligence Project, Japan\\
$^4$Department of Computer Science, Illinois Institute of Technology, USA\\
\emails
\{xuanhanyu, wu\_zhiliang, csjyang\}@njust.edu.cn,
\{yihong.xu, xavier.alameda-pineda\}@inria.fr, \\
shuo.chen.ya@riken.jp,
yyan34@iit.edu
}

\begin{document}

\maketitle

\begin{abstract}
The recent success of audio-visual representation learning 
can be largely attributed to their pervasive property of audio-visual synchronization, 
which can be used as self-annotated supervision. 
As a state-of-the-art solution, 
\emph{Audio-Visual Instance Discrimination} (\emph{AVID}) extends instance discrimination to the audio-visual realm.
Existing \emph{AVID} methods construct the contrastive set by random sampling
based on the assumption that the audio and visual clips from all other videos are not semantically related. We argue that this assumption is rough, since the resulting contrastive sets have a large number of faulty negatives.
In this paper, we overcome this limitation by proposing a novel Active Contrastive Set Mining (\emph{ACSM}) that aims to mine the contrastive sets with informative and diverse negatives for robust \emph{AVID}.
Moreover, we also integrate a semantically-aware hard-sample mining strategy into our \emph{ACSM}.
The proposed \emph{ACSM} is implemented into two most recent state-of-the-art \emph{AVID} methods and significantly improves their performance.
Extensive experiments conducted on both action and  sound recognition on multiple datasets show the remarkably improved performance of our method.
\end{abstract}

\section{Introduction}

Visual scenes are typically synchronised with a mixture of sounds, for instance moving lips and speech or moving cars and engine noise\cite{xuan2020cross}.
This pervasive audio-visual synchronization comes from the fact that sound is produced by the vibration of objects.
Through such an audio-visual synchronised perception~\cite{smith2005development}, 
humans are capable of developing skills to better perceive the world.
As for machine intelligence, audio-visual synchronization in videos also 
raises the possibility to develop similar skills by investigating audio-visual representation learning.
In addition,
different from expensive and subjective manual annotations, audio-visual synchronization provides free and extensive self-annotated supervisions for exploring audio-visual representation learning using numerous videos available on the Internet.

Recent studies on self-supervised audio-visual representation learning
\cite{arandjelovic2017look,arandjelovic2018objects,korbar2018cooperative,owens2018audio} set up a binary verification problem that predicts whether an input audio-visual pair is correct, that is an ``in-sync'' video and audio, or incorrect, that is constructed by using ``out-of-sync'' audio~\cite{korbar2018cooperative} or audio from a different video~\cite{arandjelovic2017look}.
However, exploiting one single pair at a time
loses the key opportunity to reason about the data distribution in large-scale.

\begin{figure}[t]
\centering
\includegraphics[width=8.2cm]{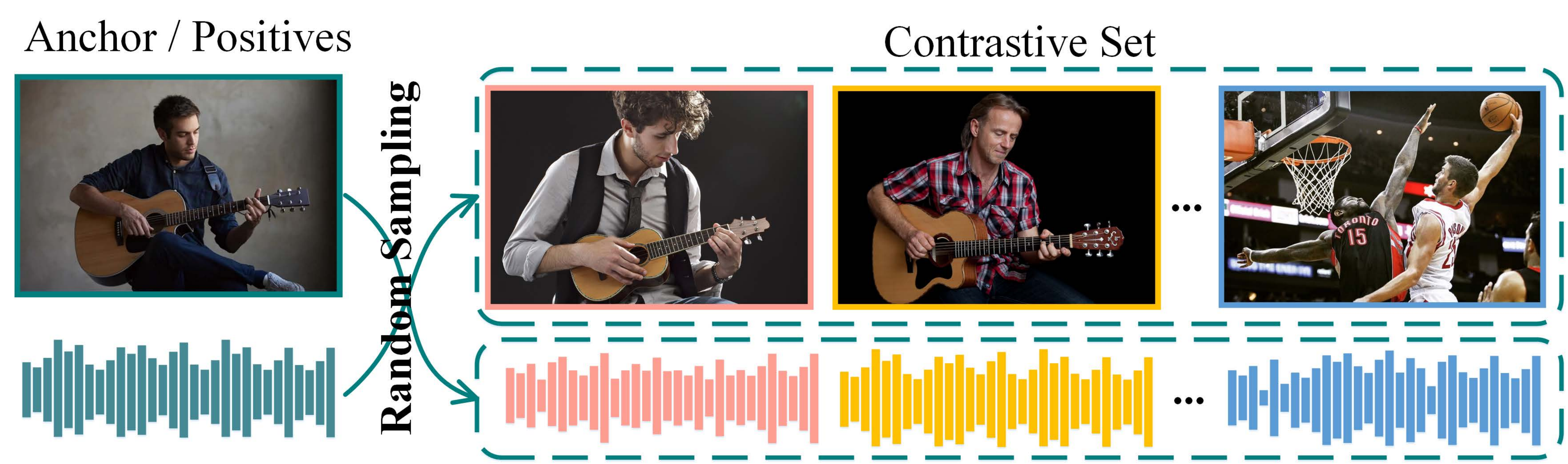}
\caption{Illustration of a contrastive set constructed by randomly sampling
in \emph{AVID}.
Such a random sampling relies on one key assumption:
the audio and visual clips from all other videos are not semantically related.
The contrastive set with such \emph{faulty negatives} undermines the primary goal of \emph{AVID},
i.e.\
we cannot keep the visual clips with guitar playing and the audio with guitar strumming both far away from and close to each other in the representation space.
}
\label{Fig1}
\end{figure}

To alleviate the above issue,
some works have attempted to improve upon the binary verification problem
by posing it as an Instance Discrimination (\emph{ID}) \cite{jaiswal2021survey} task,
called Audio-Visual Instance Discrimination (\emph{AVID}).
Specifically,
\cite{morgado2021robust,morgado2021audio}
propose a cross-modal version of \emph{ID}
and \cite{ma2021active} present a multi-modal extension of Momentum Contrast (MoCo) \cite{he2020momentum}.
They have achieved impressive performance on \emph{AVID}
through constructing a contrastive set with large negatives by randomly sampling 
from a memory bank \cite{wu2018unsupervised} or a queue-based dictionary \cite{he2020momentum}.
The reasons for the success are two-fold:
on one hand,
the introduction of a contrastive set rather than a single pair facilitates reasoning about the data distribution.
On the other hand,
memory bank and dictionary
decouple the size of the contrastive set from the mini-batch size, allowing it to be large. 
Theoretically,
a large contrastive set can achieve a tighter lower bound on Mutual Information (\emph{MI})
\cite{mcallester2020formal}.

However,
these methods rely on one key assumption when constructing the contrastive set for \emph{AVID}:
the audio and visual clips from all other videos are not semantically related.
In practice, such an assumption does not hold for a significant amount of real-world videos.
Simply increasing the size of the contrastive set beyond a threshold does not improve
or even harm the performance of learned representations on downstream tasks \cite{pmlr2019saunshi}.
We argue that such a contrastive set,
constructed by randomly sampling rather than carefully designing,
is much responsible for this:
as shown in Fig.\ref{Fig1},
it can lead to \emph{faulty negatives} (semantically related to the anchor).
The contrastive set with such faulty negatives undermines the primary goal of \emph{AVID},
i.e.,
we cannot keep the audio-visual pairs with related semantics both far from and close  to each other in the representation space.
Also, this issue can get aggravated
as the size of the contrastive set increases.

For this purpose,
we propose a method called Active Contrastive Set Mining (\emph{ACSM})
that aims to mine the contrastive sets
with informative and diverse negatives for robust \emph{AVID}.
Specifically,
the semantic structures are explicitly imposed to unlabelled audio-visual clips
to encourage learning a ``semantics-aware'' discriminative space.
At this point,
the contrastive set can be constructed under 
the guidance of progressive semantics during the training of our \emph{ACSM}.
Different from the existing  \emph{AVID}
\cite{morgado2021robust,morgado2021audio,ma2021active}  
which pull away each audio or visual clips from all other videos in the representation space,
our~\emph{ACSM} takes also the diverse semantics into account, 
i.e., only the audio-visual pairs with diverse semantics are far away from each other.
Moreover,
benefiting from our semantics-aware beyond instance-specific~\emph{AVID},
we introduce semantic ambiguity,
rather than simple time gaps \cite{Bruno2018Cooperative}, to mine hard samples,
which can be easily integrated into~\emph{AVID}. In summary, the main contributions of this work are re-emphasised as follows:
\begin{itemize}
    \item An active contrastive set mining (\emph{ACSM}) is proposed for mining the contrastive sets with informative and diverse negatives for robust \emph{AVID};
    \item We elegantly  extend our \emph{ACSM} to two state-of-the-art \emph{AVID} methods, i.e., Cross-modal ID \cite{ma2021active} and Multi-modal MoCo \cite{morgado2021robust,morgado2021audio}, which then achieves significant improvements;
    \item With the semantics from our ACSM, a hard sample mining strategy based on the semantic ambiguity is introduced to boost  discriminative ability of \emph{AVID}.
\end{itemize}

Extensive experiments and remarkable performance gains show the effectiveness of our method.
Our method also achieves state-of-the-art transfer learning performance,
both action recognition on UCF101 and HMDB51 datasets
and sound recognition on ESC-50 and DCASE datasets,
when pre-training on the subset of Audioset dataset.

\section{Related Work}

\paragraph{Instance Discrimination}~conducts contrastive learning 
by treating each sample as a separate class
and pulling the positives (``\emph{similar}'') closer 
while pushing the negatives (``\emph{dissimilar}'') far away 
to learn \emph{instance-specific} discriminative representation.
MoCo \cite{he2020momentum} 
adopts an online encoder and a momentum encoder to receive two views of a sample as positive pairs.
A queue-based dictionary is built to store negative samples. 
SimCLR \cite{chen2020simple} 
pushes self-supervised pre-training to comparable effects as the supervised methods. 
They also carefully sort out the tricks for effective training, 
such as longer training time or stronger data augmentation. 
\cite{chen2020improved} proposes MoCo-v2 
and refreshes the performance of self-supervised learning once again.
BYOL \cite{susmelj2020lightly} adds a predictor to learn the map from the online
encoder to the momentum encoder instead of displaying positive
samples. 
Through the stop gradient mechanism, they skillfully discard the negatives.
Simsiam \cite{chen2021exploring}  lets the target encoder and the online encoder be the same and points out that the predictor and the stop gradient mechanism are sufficient conditions
for self-supervised training.

These works rely heavily on proper data augmentation setups.
Instead, we focus on cross-modal instance discrimination,
which avoids this issue through the inherent and pervasive synchronization between audio-visual messages.
Furthermore,
our goal is to be ``\emph{semantics-aware}'', not just instance-specific instance discrimination
by introducing the idea of deep clustering \cite{zhan2020online}.

\paragraph{Self-Supervised Audio-Visual Representation Learning.}
Besides computer vision,
several approaches attempt to learn audio-visual representations
in a self-supervised manner.
\cite{arandjelovic2017look,arandjelovic2018objects}
propose to learn audio-visual representations by solving a binary verification problem
that predicts whether the input audio-visual pair is matched or not.
\cite{korbar2018cooperative,owens2018audio}
predict if audio-visual clips are synchronized in time,
and \cite{morgado2020learning} predicts if the audio-visual clips extracted from the $360^\circ$ videos are aligned in space.
\cite{morgado2021robust,morgado2021audio,ma2021active}
improve upon the binary verification problem
by posing it as an instance discrimination task,
where the size of the contrastive set is decoupled from the mini-batch size, allowing it to be large.
In this paper,
we explore an online mining strategy, rather than random sampling, to construct a reliable contrastive set.

\section{Audio-Visual Instance Discrimination}

\begin{figure}[htbp]
\centering
\includegraphics[width=8.2cm]{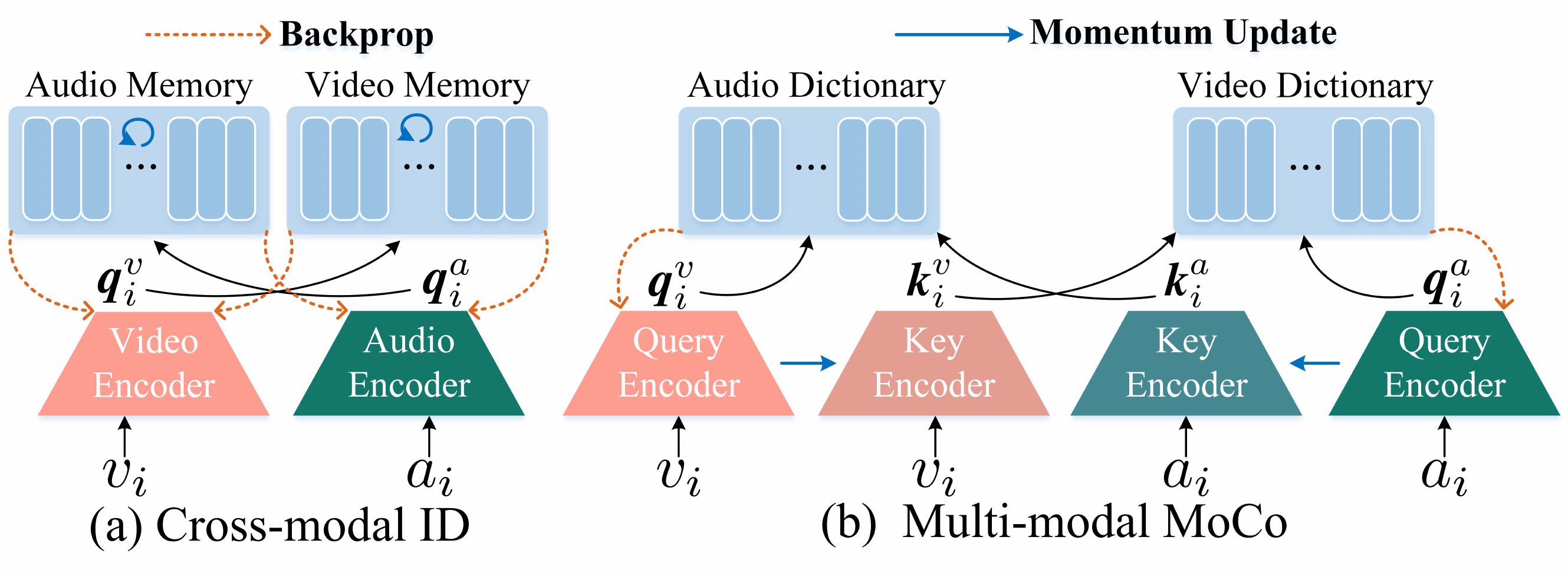}
\caption{
The diagram of Cross-modal ID \protect\cite{morgado2021robust,morgado2021audio}
and Multi-modal MoCo \protect\cite{ma2021active},
where $(a_i,v_i)$ denotes the input audio-visual pair.
We use $f(\cdot)$ indicates the \emph{query} encoder with learnable parameters $\bm{\theta}$,
updated by the gradients,
and $h(\cdot)$ represents the \emph{key} encoder with learnable parameters $\bm{\delta}$,
updated by the momentum.
The \emph{superscript} is used  to distinguish various data streams from audio or visual modality,
where the query encoder and key encoder from the same modality
have the same structure but independent parameters.
Formally,
the formulation can be expressed as:
$\textbf{\emph{q}}_i^v = f^v(v_i ; \bm{\theta}^v)$,
$\textbf{\emph{q}}_i^a = f^a(a_i ; \bm{\theta}^a)$,
$\textbf{\emph{k}}_i^v = h^v(v_i ; \bm{\delta}^v)$ and
$\textbf{\emph{k}}_i^a = h^a(a_i ; \bm{\delta}^a)$.
}
\label{Fig2}
\end{figure}

Let $(A,V)$ be a collection of $N$ videos,
where $A= \{ a_i\}_{i=1}^{N}$, $V= \{ v_i\}_{i=1}^{N}$,
each audio-visual pair $\{ a_i,v_i \}$ is from the same block of a video.
Here, $a_i \in \mathbb{R}^{ H_a \times W_a}$
refers to the spectrogram of the raw audio waveform,
where $H_a$ is time step and $W_a$ is frequency band.
$v_i \in \mathbb{R}^{F \times 3 \times H_v \times W_v}$
is the visual clip containing $F$ RGB frames
with a height of $H_v$ and a width of $W_v$.
The learning objective of \emph{AVID}
is to encourage the representations of audio-visual clips to be similar
if they come from the same temporal block of a video and vice-versa.

Formally, \emph{AVID} employs a Noise-Contrastive Estimation (\emph{NCE}) \cite{gutmann2010noise}
to learn two encoders independently,
$f (\cdot;\bm{\theta})$ and
$h (\cdot;\bm{\delta})$,
by matching $\textbf{\emph{q}}_i$ with $\textbf{\emph{k}}_i$ against its contrastive set
$ \mathcal{K}_{i} = \{ \textbf{\emph{k}}_j^i \}_{j=1}^K $:
\begin{equation}
\mathcal{L}_{nce}(\textbf{\emph{q}}_i; \textbf{\emph{k}}_i, \mathcal{K}_i)=
-\log
\frac{\exp[\cos(\textbf{\emph{q}}_i,\textbf{\emph{k}}_i)/ \tau ]}
{\sum_{\tilde{\textbf{\emph{k}}}} \exp[\cos(\textbf{\emph{q}}_i,\tilde{\textbf{\emph{k}}})/ \tau ]},
\label{Eq1}
\end{equation}
where
$\textbf{\emph{q}}_i = f (\cdot)$ and
$\textbf{\emph{k}}_i = h (\cdot)$ are the outputs of
query encoder $f$ and key encoder $h$
with parameters $\bm \theta$ and $\bm \delta$ respectively,
$\tilde{\textbf{\emph{k}}} \in \mathcal{K}_i \cup \{ \textbf{\emph{k}}_i\}$,
$\cos(\cdot,\cdot)$
is the cosine similarity between a pair of representations,
$\tau$ is the temperature to control the concentration degree of distribution
and $K$ is the size of the contrastive set.

Minimising \emph{NCE} loss means maximising a lower bound on the \emph{MI} between audio-visual pairs,
which allows the model to learn to discriminate
whether audio-visual pairs are ``similar'' or ``dissimilar''.
The final objective $\mathcal{L}_{nce}$ of \emph{AVID} consists of two terms:
\begin{equation}
\mathcal{L}_{nce} =
\mathcal{L}_{nce}(\textbf{\emph{q}}_i^v; \textbf{\emph{k}}_i^a, \mathcal{K}_i^a) +
\mathcal{L}_{nce}(\textbf{\emph{q}}_i^a; \textbf{\emph{k}}_i^v, \mathcal{K}_i^v),
\label{avID-NCE}
\end{equation}
where the former indicates visual-to-audio component
and the latter refers to audio-to-visual component.

There are two most recently proposed approaches
for performing \emph{AVID},
i.e., Cross-modal ID \cite{morgado2021robust,morgado2021audio}
and Multi-modal MoCo \cite{ma2021active}.
The diagram of these two approaches is depicted in Fig.\ref{Fig2}.
We make a brief summary of the main differences and connections between the two.

\paragraph{Momentum Update.}
As shown in Fig.\ref{Fig2},
both Cross-modal ID and Multi-modal MoCo
employ a momentum update
to maintain consistency during training:
\begin{equation}
\textbf{\emph{u}}_{_{old}} \leftarrow m \textbf{\emph{u}}_{_{old}} + (1-m) \textbf{\emph{u}}_{_{new}},
\label{Eq3}
\end{equation}
where $m \in [0,1)$ is a momentum coefficient,
$\textbf{\emph{u}}$ is the content to be updated.
The former utilizes a momentum-based moving average memory bank \cite{wu2018unsupervised},
while the latter exploits a momentum-based slowly-changing encoder.
Besides,
Multi-modal MoCo performs a cross-modal online dictionary look-up,
i.e., the encoded representations of the current mini-batch are enqueued
and the oldest ones are dequeued.

Given a query,
its contrastive set $\mathcal{K}_i$ in Eq.\ref{Eq1}
is constructed by randomly sampling from a semantically indistinguishable memory bank or queue-based dictionary
as there are no annotations available.
Such a random sampling strategy makes the contrastive set contain massive faulty negatives
and undermines the primary goal of \emph{AVID},
resulting in no improvement or even impairment
when simply increasing the size of the contrastive set.

\section{Active Contrastive Set Mining}
We believe that the random sampling without semantic guidance is the main cause of the above problem
when constructing the contrastive set.
For this purpose,
we introduce a novel semantic-sensitive library\footnote{
We use \emph{library} to collectively denote \emph{memory} bank and \emph{dictionary}.
That is to say, the library can be updated in a queue-based or momentum-based manner.}
for mining the contrastive set with informative and diverse negatives for robust \emph{AVID}.
Such libraries serve three functions:
\begin{itemize}
    \item playing the role of conventional memory bank or dictionary
to store a series of representations on-the-fly;
    \item imposing the underlying semantic structures on the audio and visual samples;
    \item providing the semantic guidance when constructing the contrastive set.
\end{itemize}

To be concrete, different from a single library with semantic mixtures,
we manage $C$ independent libraries
$\mathcal{M}=\{ M_m \}_{m=1}^C$,
where each library $M_m$ corresponding to \emph{one semantics}.
They are constructed by a learnable classifier that predicts the pseudo-label of a query.
For a query $\textbf{\emph{q}}_i$ with pseudo-label $y_i$,
we can construct its contrastive set $\mathcal{K}_{i}$
in a manner similar to supervised contrastive learning \cite{khosla2020supervised}:
\begin{equation}
\mathcal{K}_{i} = \{ \hat{\textbf{\emph{k}}_j} ~|~ \hat{\textbf{\emph{k}}_j} \in M_m,
~ \forall~ m \neq y_i \},
\label{ContrastiveSet}
\end{equation}
where $j \in [1,K]$ and $m, y_i \in [1,C]$.
After every backward pass,
its key $\textbf{\emph{k}}_i$
is used to update the corresponding $M_{y_i}$ in a queue-based or momentum-based manner.

\begin{table*}[t]
\caption{Top-1 accuracy of action recognition on UCF-101 and HMDB-51 datasets.}
\centering
\begin{tabu}{c|c|c|c|c|c}
\tabucline[1pt]{-}
    \textbf{Methods}    &
    \makecell[c]{\textbf{Pretraining} \\ \textbf{Dataset}}  &
    \makecell[c]{\textbf{Finetune} \\ \textbf{Resolution}}  &
    \textbf{Architecture} &
    \textbf{UCF-101}   &
    \textbf{HMDB-51}\\
\tabucline[1pt]{-}

    Shuffle\&Learn \cite{misra2016shuffle}   &
    \multirow{4}{*}{UCF-101} &
    \multirow{4}{*}{$1 \times 227^2$} &
    \multirow{2}{*}{CaffeNet} &
    $50.2$    &
    $18.1$    \\
    \cline{1-1} \cline{5-6}

    OPN \cite{lee2017unsupervised}         &  &  &                     &  $56.3$    &  $23.8$     \\
    \cline{1-1} \cline{4-6}

    ST-Order \cite{buchler2018improving}    &  &  &         VGG         &  $58.6$    &  $25.0$     \\
    \cline{1-1} \cline{4-6}

    CMC \cite{tian2020contrastive}         &  &  &     3D-ResNet18     &  $59.1$    &  $26.7$     \\
\tabucline[1pt]{-}

    CBT \cite{Sun2019ContrastiveBT}       &  Kinetics-600   & \multirow{3}{*}{$16 \times 122^2$} &  {S3D \& BERT} &  $79.5$  &  	$44.6$    \\
    \cline{1-2} \cline{4-6}

    3D-RotNet \cite{jing2018self} & \multirow{13}{*}{Kinetics-400} &  &  3D-ResNet18    &  $62.9$ &   $33.7$   \\
    \cline{1-1} \cline{4-6}

    ClipOrder \cite{xu2019self}       &  &  &  R(2+1)D-18    &  $72.4$    &  	$30.9$  \\
    \cline{1-1} \cline{3-6}

    DPC \cite{han2019video}             &  &  $25 \times 128^2$ &  3D-ResNet34    &  $75.7$   &   $35.7$    \\
    \cline{1-1} \cline{3-6}

    $L^3$-Net \cite{arandjelovic2017look}              &  &  $16 \times 224^2$ &  2D-Conv         &  $74.4$   &   $47.8$    \\
    \cline{1-1} \cline{3-6}

    AVTS \cite{korbar2018cooperative}            &  &  $25 \times 224^2$ &  MC3            &  $85.8$   &   $56.9$    \\
    \cline{1-1} \cline{3-6}

    XDC \cite{alwassel2019self}             &  & \multirow{8}{*}{$8 \times 224^2$} & \multirow{7}{*}{R(2+1)D-18} &
        $74.2$    &   $39.0$  \\
    \cline{1-1} \cline{5-6}

    xID \cite{morgado2021audio}             &  &  &  &
        $82.3$    &  $49.1$ \\
    \cline{1-1} \cline{5-6}

    Robust-xID \cite{morgado2021robust}      &  &  &  &
        $81.9$    &  $49.5$ \\
    \cline{1-1} \cline{5-6}

    xID + CMA \cite{morgado2021audio}      &  &  &  &
        $83.7$    &  $50.2$ \\
    \cline{1-1} \cline{5-6}

    MmMoCo \cite{ma2021active}         &  &  &  &
        $83.5$ &  $49.4 $\\
    \cline{1-1} \cline{5-6}

    MmMoCo + AS \cite{ma2021active}        &  &  &  &
         $85.6$&  $58.6 $   \\
    \cline{1-1} \cline{5-6}

    xID + \textbf{ACSM}     (\textbf{\emph{Ours}})  &  &  &  &
        $\bm{86.7}$&  $\bm{60.4}$ \\
    \cline{1-1} \cline{5-6}

    MmMoCo + \textbf{ACSM}  (\emph{\textbf{Ours}})  &  &  &  &
        $\bm{88.4}$&  $\bm{61.3}$ \\
\tabucline[1pt]{-}
\end{tabu}
\label{ActionR}
\end{table*}

The samples with various pseudo-labels are put in separate libraries.
As a result,
one library can be viewed as one semantic ``\emph{cluster}'',
where the samples in same library
can be taken as anchors to describe this semantics.
These semantic libraries can naturally serve as the semantic decision boundaries
based on sample-anchor similarity:
\begin{equation}
g_{ij} =
\frac
{\sum_{\textbf{\emph{m}} \in M_j}
\exp[\cos(\textbf{\emph{q}}_i,\textbf{\emph{m}}) / \tau]}
{\sum_{m=1}^C
\sum_{\textbf{\emph{m}} \in M_m} \exp[\cos(\textbf{\emph{q}}_i, \textbf{\emph{m}}) / \tau]},
\label{Eq5}
\end{equation}
Given a query $\textbf{\emph{q}}_i$
its similarity-based semantics assignments are represented as
$\textbf{\emph{g}}_i = [g_{i1},\cdots,g_{ij},\cdots,g_{iC}]$,
where $\textbf{\emph{g}}_i \in \mathbb{R}^C$.
With such potential memberships,
we can train the classifier 
which aims to minimise $\textbf{\emph{g}}_i$ 
and the semantics predictions $\textbf{\emph{p}}_i$,
where $\textbf{\emph{p}}_i$ is yielded by
the classifier $r(\cdot ; \bm \gamma)$
with the learnable parameters $\bm \gamma$,
i.e., $\textbf{\emph{p}}_i = r(\textbf{\emph{q}}_i)$.
We propagate the gradient back to the classifier only
to avoid feature learning from unreliable semantics.
With the updated classifier $r(\cdot)$,
we update the predicted semantics $y_i$
in a maximum likelihood manner for the contrastive set construction in Eq.~\ref{ContrastiveSet}.
At the same time,
the samples are assigned to the library with the most similar anchors
and each library holds its own attribute
that makes it different from the others.

\paragraph{Discussion.}
Intuitively,
at the beginning of training,
the weak classifier provides near-random semantics predictions,
so that our semantic libraries contain no semantic structures.
At this point,
the contrastive set constructed by Eq.\ref{ContrastiveSet} approximates to a random sampling.
With the progressive classifier,
our semantic libraries cover increasingly clear semantic structures
and provide effective semantic guidance when constructing the contrastive set,
allowing our contrastive set to contain more informative and diverse negatives.

Furthermore,
both Cross-modal ID \cite{morgado2021robust,morgado2021audio}
and Multi-modal MoCo \cite{ma2021active}
only optimise instance-specific discriminative representations,
leading them to be not semantics sensitive
and unaware of any potential non-linear intra-semantics variation.
Compared with them,
our \emph{ACSM} maximises not only the instance-level diversity within the library,
but also semantic-level compactness among libraries.
The samples belonging to the same semantic library may share a common contrastive set.
It means that the samples within the library
are indirectly pushed closer in the representation space,
resulting in more compact within-library representations.

\paragraph{Hard Sample Mining.}
The samples that are easily discriminated by the current model 
tend to contribute less to instance discrimination \cite{sheng2020mining}.
Hard sample mining is important in the instance discrimination field
but is especially essential in the audio-visual field.
From the information-theoretic perspective \cite{chen2018learning},
audio-visual messages in videos contain higher \emph{MI}
due to the higher dimensionality
(i.e., temporal and multi-modal).
The instances with higher \emph{MI} mean that they are easier to discriminate.
For this purpose,
\cite{Bruno2018Cooperative} perform the hard sample mining
based on one key assumption:
the smaller the time gap is between audio-visual clips of the same video,
the harder it is to discriminate them.
Instead of making such a assumption,
we mine hard samples by emphasising the samples with ambiguous semantics.
Given a query $\textbf{\emph{q}}_i$ with pseudo-label $y_i^{e}$ at $e$th epoch,
its semantic ambiguity is denoted as:
\begin{equation}
s_i^e = s_i^{e-1} + 1, \text{~if~} y_i^{e} \neq y_i^{e-1}
\label{Eq6}
\end{equation}
Those samples that are frequently swapped across semantics
will be assigned higher weights in Eq.\ref{avID-NCE} to provide more useful discriminative clues.
Overall, we apply our ACSM to
Cross-modal ID \cite{ma2021active} and
Multi-modal MoCo \cite{morgado2021robust,morgado2021audio}.

\begin{table*}[t]
\caption{Top-1 accuracy of sound recognition on ESC-50 and DCASE datasets.
}
\centering
\begin{tabu}{c|c|c|c|c}

\tabucline[1pt]{-}
    \textbf{Methods} &
    \makecell[c]{\textbf{Pretraining} \\ \textbf{Dataset}}  &
    \textbf{Architecture}    &
    \textbf{ESC-50}  &
    \textbf{DCASE}   \\
\tabucline[1pt]{-}

    Random Forest \cite{piczak2015esc}   &
    \multirow{3}{*}{ESC-50} &  MLP   & $44.3$ & $-$ \\
    \cline{1-1} \cline{3-5}

    Piczak ConvNet \cite{piczak2015environmental}  &  &
    \multirow{2}{*}{ConvNet-4} & $64.5$ & $-$ \\
    \cline{1-1} \cline{4-5}

    ConvRBM \cite{sailor2017unsupervised}  &  &  & $86.5$ & $-$ \\
\tabucline[1pt]{-}

    Sound-Net \cite{aytar2016soundnet}    &
    \multirow{2}{*}{SoundNet} & \multirow{2}{*}{ConvNet-8} & $74.2$ & $88$ \\
    \cline{1-1} \cline{4-5}

    $L^3$-Net \cite{arandjelovic2017look}  &  &  & $79.3$ &  $93$ \\
\tabucline[1pt]{-}

    AVTS \cite{korbar2018cooperative}   &
    \multirow{9}{*}{Kinetics-400} &  VGG-8    &  $76.7$ &  $91$\\
    \cline{1-1} \cline{3-5}

    XDC \cite{alwassel2019self} &  & ResNet-18 & $78.5$ &  $-$ \\
    \cline{1-1} \cline{3-5}

    xID \cite{morgado2021audio} &  & \multirow{7}{*}{ConvNet-9} &  $77.6$ &  $93$ \\
    \cline{1-1} \cline{4-5}

    xID + CMA \cite{morgado2021audio}  &  &  & $79.1$ &  $93$   \\
    \cline{1-1} \cline{4-5}

    Robust-xID \cite{morgado2021robust} &  &  &  $80.2$&  $92.4$    \\
    \cline{1-1} \cline{4-5}

    MmMoCo \cite{ma2021active}   &  &  &  $78.8$ &  $93.6$\\
    \cline{1-1} \cline{4-5}

    MmMoCo + AS \cite{ma2021active} &  &  &  $86.4$ &  $94.3$ \\
    \cline{1-1} \cline{4-5}

    xID + \textbf{ACSM} (\textbf{\emph{Ours}})   &  &  &  $\bm{88.5}$ &   $\bm{95.2}$\\
    \cline{1-1} \cline{4-5}

    MmMoCo + \textbf{ACSM} (\textbf{\emph{Ours}})    &  &  &  $\bm{89.3}$ &  $\bm{94.7}$ \\
\tabucline[1pt]{-}
\end{tabu}
\label{SoundR}
\end{table*}
\begin{table*}[t]
\caption{Action recognition accuracy of linear probing on Kinetics
and sound recognition accuracy of linear probing on ESC-50,
where the evaluated representations come from various blocks of the encoder,
$\color{teal} \uparrow$ indicates the increase in accuracy.}
\centering
\begin{tabu}{c|cccc|cccc}
\tabucline[1pt]{-}
\rowcolor[HTML]{FFFFFF}
\cellcolor[HTML]{FFFFFF} &
\multicolumn{4}{c|}{\cellcolor[HTML]{FFFFFF}\textbf{Kinetics-400}} &
\multicolumn{4}{c }{\cellcolor[HTML]{FFFFFF}\textbf{ESC-50}} \\

\rowcolor[HTML]{FFFFFF}
\multirow{-2}{*}{\cellcolor[HTML]{FFFFFF}\textbf{Method}}  &
\multicolumn{1}{c }{\cellcolor[HTML]{FFFFFF}\emph{block1}} &
\multicolumn{1}{c }{\cellcolor[HTML]{FFFFFF}\emph{block2}} &
\multicolumn{1}{c }{\cellcolor[HTML]{FFFFFF}\emph{block3}} &
\emph{block4} &
\multicolumn{1}{c }{\cellcolor[HTML]{FFFFFF}\emph{block1}} &
\multicolumn{1}{c }{\cellcolor[HTML]{FFFFFF}\emph{block2}} &
\multicolumn{1}{c }{\cellcolor[HTML]{FFFFFF}\emph{block3}} &  \emph{block4} \\
\tabucline[1pt]{-}

\rowcolor[HTML]{EFEFEF} xID &
\multicolumn{1}{c }{\cellcolor[HTML]{EFEFEF}$19.80$} &
\multicolumn{1}{c }{\cellcolor[HTML]{EFEFEF}$26.98$} &
\multicolumn{1}{c }{\cellcolor[HTML]{EFEFEF}$34.81$} &
$39.95$ &
\multicolumn{1}{c }{\cellcolor[HTML]{EFEFEF}$67.25$} &
\multicolumn{1}{c }{\cellcolor[HTML]{EFEFEF}$73.15$} &
\multicolumn{1}{c }{\cellcolor[HTML]{EFEFEF}$74.80$} & $75.05$ \\

\rowcolor[HTML]{FFFFFF} xID + \textbf{ACSM} &
\multicolumn{1}{c }{\cellcolor[HTML]{FFFFFF}$\color{teal} 1.26 \uparrow$} &
\multicolumn{1}{c }{\cellcolor[HTML]{FFFFFF}$\color{teal} 1.69 \uparrow$} &
\multicolumn{1}{c }{\cellcolor[HTML]{FFFFFF}$\color{teal} 2.37 \uparrow$} &
$\color{teal} 3.56 \uparrow$ &
\multicolumn{1}{c }{\cellcolor[HTML]{FFFFFF}$\color{teal} 3.85 \uparrow$} &
\multicolumn{1}{c }{\cellcolor[HTML]{FFFFFF}$\color{teal} 5.56 \uparrow$} &
\multicolumn{1}{c }{\cellcolor[HTML]{FFFFFF}$\color{teal} 7.12 \uparrow$} &
$\color{teal} 8.71 \uparrow$    \\

\rowcolor[HTML]{EFEFEF} MmMoCo &
\multicolumn{1}{c }{\cellcolor[HTML]{EFEFEF}$20.12$} &
\multicolumn{1}{c }{\cellcolor[HTML]{EFEFEF}$26.48$} &
\multicolumn{1}{c }{\cellcolor[HTML]{EFEFEF}$35.35$} &
$39.75$ &
\multicolumn{1}{c }{\cellcolor[HTML]{EFEFEF}$66.86$} &
\multicolumn{1}{c }{\cellcolor[HTML]{EFEFEF}$72.69$} &
\multicolumn{1}{c }{\cellcolor[HTML]{EFEFEF}$74.93$} & $76.56$  \\

\rowcolor[HTML]{FFFFFF} MmMoCo + \textbf{ACSM} &
\multicolumn{1}{c }{\cellcolor[HTML]{FFFFFF}$\color{teal} 0.94 \uparrow$} &
\multicolumn{1}{c }{\cellcolor[HTML]{FFFFFF}$\color{teal} 1.76 \uparrow$} &
\multicolumn{1}{c }{\cellcolor[HTML]{FFFFFF}$\color{teal} 2.68 \uparrow$} &
$\color{teal} 3.35 \uparrow$    &
\multicolumn{1}{c }{\cellcolor[HTML]{FFFFFF}$\color{teal} 4.11 \uparrow$} &
\multicolumn{1}{c }{\cellcolor[HTML]{FFFFFF}$\color{teal} 5.98 \uparrow$} &
\multicolumn{1}{c }{\cellcolor[HTML]{FFFFFF}$\color{teal} 6.94 \uparrow$} &
$\color{teal} 8.58 \uparrow$    \\
\tabucline[1pt]{-}

\end{tabu}
\label{Tab_blocks}
\end{table*}

\section{Experiments}
For quantitative evaluation,
we evaluate both the pre-trained  visual and audio representations using transfer learning.
Concretely,
these pre-trained representations are evaluated by linear probing,
i.e.,
keep the pre-trained network fixed and train linear classifiers.

\subsection{Experimental Setups}

\paragraph{Data preprocessing.}
Video clips are extracted at a frame rate of $16$fps
with some standard data augmentation strategies,
i.e., random multi-scale cropping,
random horizontal flipping,
color jittering and temporal jittering.

For audio clips,
each audio with a duration of $2$s
is randomly sampled within $0.5$s at a $24$kHz sampling rate.
Then, we convert the spectrogram of the audio track to a log scale,
its intensity is $Z$-normalized using mean and standard deviation values obtained on the training set.
Volume jittering and temporal jittering are used for audio augmentation.

\paragraph{Video and audio encoder.}
We employ the  R(2+1)D network \cite{tran2018closer}
with $18$ layers (Sect.\ref{Sec42}) or $9$ layers (Sect.\ref{Sec43})
as the video encoder
and a 9-layer 2D CNN with batch normalization as the audio encoder.
The outputs of the video encoder and audio encoder are max-pooled,
projected into $128$ dimensions using an MLP,
and then normalized into the unit sphere,
where the MLP is composed of three FC layers with $512$ hidden units.
We utilize the $128$-dimensional representation to update our semantic libraries.

\paragraph{Implementation details.}
For fair comparisons,
we follow the same settings of \cite{morgado2021audio}.
Respectively,
the input of video encoder is set as
$8$ frames of size $224 \times 224$  (Sect.\ref{Sec42})
and $16$ frames of size $122 \times 122$  (Sect.\ref{Sec43}).
The spectrogram size is set as
$200 \times 257$ (Sect.\ref{Sec42})
and $100 \times 129$ (Sect.\ref{Sec43}).
We separately utilize Kinetics-$400$ dataset containing $240K$ videos (Sect.\ref{Sec42})
and Audioset dataset with a random sample of $100K$ videos (Sect.\ref{Sec43})
to pre-train our model.
The temperature parameter $\tau$ in Eq.\ref{Eq1} and Eq.\ref{Eq5} is set to $0.07$.
The momentum coefficient $m$ is set to $0.9$.
The number of semantic libraries $C$ is set to $50$.
The size of the contrastive set $K$ is set to $8192$.
Each semantic-sensitive library stores $K/(C-1)$ representations.
Our model is trained with the Adam optimizer for $400$ epochs with a learning rate of $1e-4$, weight decay of $1e-5$, and batch size of $256$.
The parameters of the encoders, classifiers, and semantic libraries in our model
are randomly initialised.

\subsection{Comparisons with SOTAs}
\label{Sec42}
It is impractical to give a direct comparison to all approaches
due to the enormous differences in the experimental setups,
e.g.,
using diverse encoder architectures,
pre-training on various datasets
and fine-tuning with different resolutions.
To empower solid comparisons,
we list these differences for each method in Tab.\ref{ActionR} and Tab.\ref{SoundR},
where Cross-modal ID and Multi-modal MoCo are abbreviated as xID and MmMoCo, respectively.
The methods with the best and second-best performance are indicated in boldface.

\paragraph{Visual representations.}
To evaluate the visual representations,
we compare the transfer performance of action recognition with previous self-supervised methods
on UCF-101 and HMDB-51 datasets.
We follow \cite{morgado2021audio} and report top-1 accuracy of video-level predictions
by averaging the clip-level accuracy of 10 uniformly sampled clips per test sample
in Tab.\ref{ActionR}.

\begin{figure}[t]
\centering
\includegraphics[width=7.5cm]{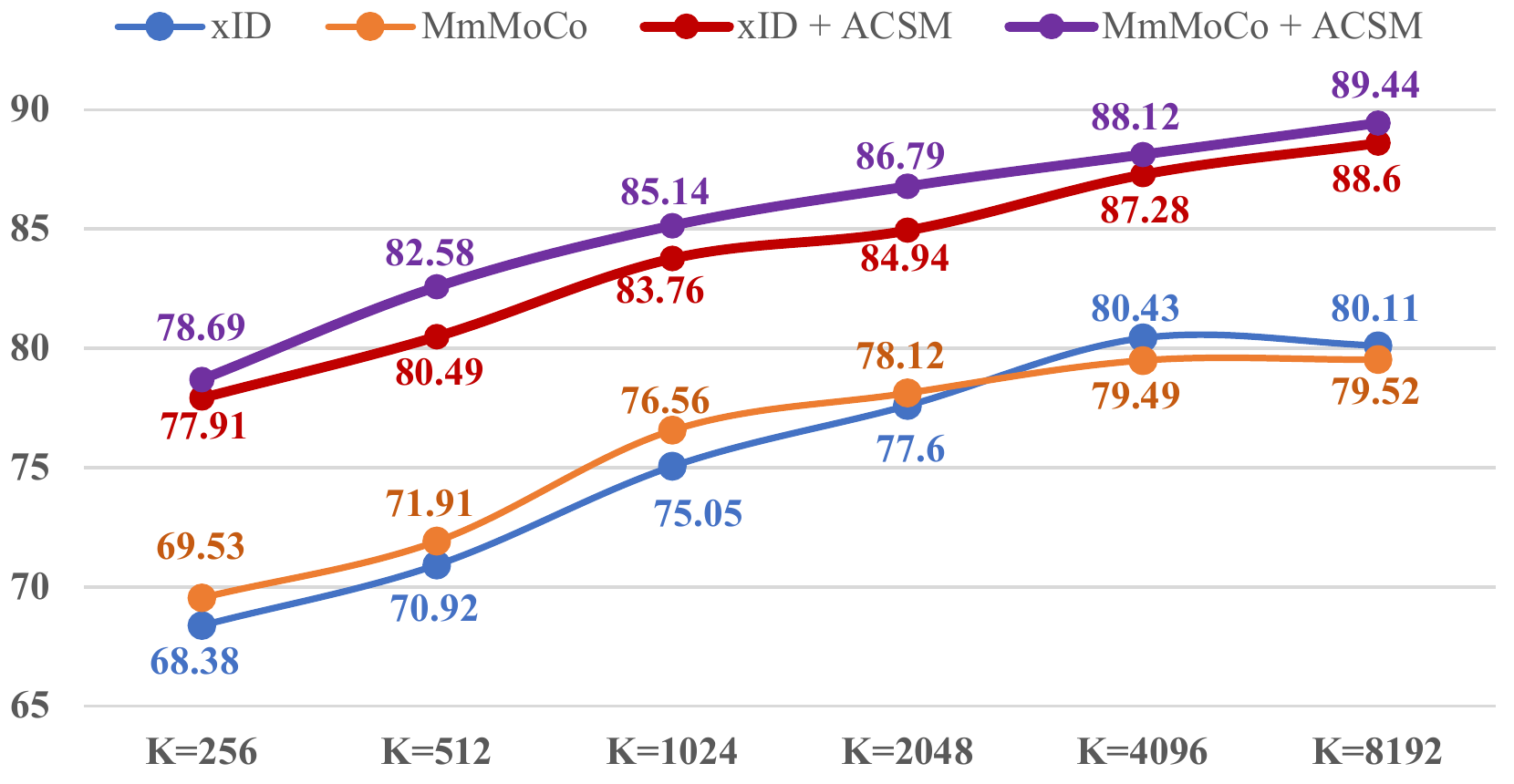}
\caption{Effects of learning with different size of contrastive set $K$.}
\label{Fig3}
\end{figure}
\begin{figure}[h]
\centering
\includegraphics[width=8.2cm]{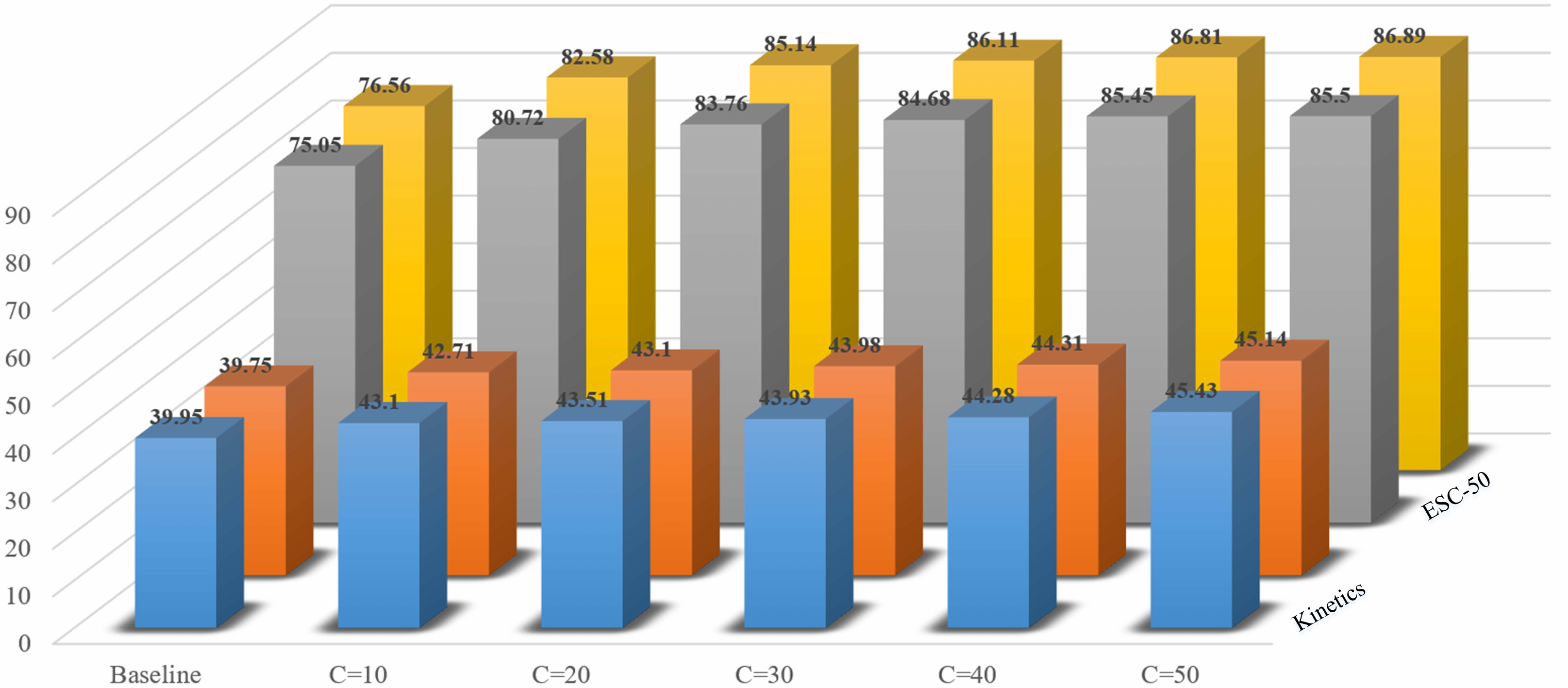}
\caption{Effects of learning with different number of semantic libraries $C$.
Both Cross-modal ID and Multi-modal MoCo are used as baselines.}
\label{Fig4}
\end{figure}

\paragraph{Audio representations.}
We evaluate the audio representations
on ESC-50 and DCASE datasets for sound recognition.
A linear one-vs-all SVM classifier is trained
on the audio representations obtained by the pre-trained models at the final layer before pooling.
We report top-1 accuracy of sample-level predictions
by averaging 10 clip-level predictions in Tab.\ref{SoundR}.

As we can see,
our approach outperforms other SOTA approaches.
Compared with MmMoCo + AS\cite{ma2021active},
the current best performing method on self-supervised audio-visual representation learning,
our model outperforms it by
$2.8\%$ on UCF101,
$2.7\%$ on HMDB51,
$2.9\%$ on ESC-50
and $0.9\%$ on DCASE.
Compared with xID \cite{morgado2021audio} and MmMoCo \cite{ma2021active}
($12$th and $15$th rows in Tab.\ref{SoundR}, $8$th and $11$th rows in Tab.\ref{ActionR}),
our ACSM achieves impressive performance improvements on both action recognition and sound recognition.

\subsection{Ablation Study}
\label{Sec43}

\paragraph{Random sampling vs. Active mining.}
We evaluate the learned visual and audio representations
in terms of the various blocks of the video and audio encoder.
Separately,
we employ the Kinetics-400 for action recognition
and ESC-50 for sound recognition.
Tab.\ref{Tab_blocks} reports the top-1 accuracy by averaging predictions over $25$ clips per video.
Introducing our ACSM in Cross-modal ID and Multi-modal MoCo,
the accuracy of both action recognition and sound recognition has been improved to varying degrees.
The increase in recognition performance
validates the benefits of a semantically sensitive library over a semantically indistinguishable one,
resulting in better generalization of the learned representations to downstream tasks.

\paragraph{Hard Sample Mining.}
We evaluate the learned visual and audio representations
according to different hard sample mining strategies.
We employ the representations from the block4 of the encoder to evaluate the performance of transfer learning on Kinetics-400 and ESC-50 datasets.
\cite{Bruno2018Cooperative} execute hard sample mining based on the time gap,
i.e., $1 \sim 200$ epochs with easy samples only and
$201 \sim 400$ epochs with $25\%$ hard samples and $75\%$ easy samples.
To make fairer comparisons,
we only execute our hard sample mining defined by Eq.\ref{Eq6} in $201 \sim 400$ epochs.
The top-1 accuracy by averaging predictions over 25 clips per video
is stated in Tab.\ref{HardMining}.
As expected,
hard sample mining improves the quality of learned representations.
Compared with \cite{Bruno2018Cooperative},
our mining strategy based on semantic ambiguity is more effective.

\paragraph{Effects of the size of the contrastive set.}
Both memory bank and queue-based dictionary decouple its size from the mini-batch size,
allowing it to be large.
Increasing the size of the contrastive set has been shown to improve the learned representations \cite{mcallester2020formal}.
We empirically study the effects of learned representations with different $K$ on ESC-50.
The results are shown in Fig.\ref{Fig3}.
As the size of the contrastive set increases,
performance almost always improves to varying degrees.
However,
the performance gains of MmMoCo gradually saturate.
the performance of xID even shows a small drop
when $K$ grows from $4096$ to $8192$.
By contrast,
our ACSM samples the contrastive set under the guidance of the semantic library,
allowing the learned representations to maintain a steady growth in accuracy.

\paragraph{Effects of the number of the semantic library.}
There is no universal principle to help determine the proper semantic library number on auxiliary tasks so to maximise their benefits.
We empirically investigate the effects of learned representations with different $C$
by using cross-modal ID and multi-modal MoCo as baselines.
The results are demonstrated in Fig.\ref{Fig4}.
All models with our semantic library can learn better representations,
resulting in higher performance on downstream tasks.
This again suggests the defects of randomly sampled contrastive set
on high-level semantic understanding.
In addition,
we also observe that the accuracy on ESC-50 grows slower and slower with the increase of semantic library number.

\begin{table}[t]
\caption{Ablation study on different strategies for hard sample mining,
where $\color{teal} \uparrow$ indicates the increase in accuracy
and \emph{on.} denotes that no hard sample mining strategy is used.}
\centering
\begin{tabu}{c|c|c|c}
\tabucline[1pt]{-}
    \textbf{Models} &
    \makecell[c]{\textbf{Different} \\ \textbf{Strategies}}&
    \textbf{Kinetics-400}&
    \textbf{ESC-50}   \\
\tabucline[1pt]{-}

\multirow{3}{*}{\textbf{xID}} &
    no. &  $39.95$   & $75.05$  \\ %
        &  time gap  & $\color{teal} 1.56 \uparrow$ &  $\color{teal} 4.67 \uparrow$ \\ %
        &  ours      & $\color{teal} 2.74 \uparrow$ &  $\color{teal} 6.72 \uparrow$ \\
\tabucline[1pt]{-}
\multirow{3}{*}{\textbf{MmMoCo}} &
    no. &  $39.75$&  $76.56$ \\ %
        &  time gap  & $\color{teal} 1.45 \uparrow$ &  $\color{teal} 4.81 \uparrow$ \\ %
        &  ours      & $\color{teal} 2.83 \uparrow$ &  $\color{teal} 6.55 \uparrow$ \\
\tabucline[1pt]{-}
\end{tabu}
\label{HardMining}
\end{table}

\section{Conclusion}
To alleviate the false negatives caused by random
sampling in \emph{AVID},
we propose an active contrastive set mining (ACSM)
that aims to learn a semantics-aware,  
not just instance-specific discriminative space.
Also, we introduce a novel hard sample mining strategy based on semantic ambiguity.
Our method can be elegantly extended to two most recent state-of-the-art AVID models
and achieves  impressive performances.

\small
\bibliographystyle{named}

\begin{spacing}{0.95}
\bibliography{ijcai22}
\end{spacing}

\end{document}